\documentstyle[aps,multicol,epsfig]{revtex}

\newcommand{\beq}{\begin{equation}}

\newcommand{\eeq}{\end{equation}}

\def\FigSize{9cm}

\begin{document}


\title{On $3+1$ Dimensional Scalar Field Cosmologies}

\author{F.L. Williams and P.G. Kevrekidis}
\address{
Department of Mathematics and Statistics, University of Massachusetts,
Amherst MA 01003-4515, USA }
\author{T. Christodoulakis, C. Helias, 
and G.O. Papadopoulos}
\address{ Nuclear and Particle Physics Section, Physics Department,
University of Athens,
Panepistimiopolis, Ilisia, Athens 15771, Greece }
\author{Th. Grammenos}
\address{Dept. of Mechanical and Industrial Engineering,
University of Thessaly,
Volos 38334, Greece }

\maketitle

\begin{abstract}
In this communication, we analyze the case of $3+1$ dimensional 
scalar field cosmologies in the presence, as well as in the
absence of spatial curvature, in isotropic, as well as in anisotropic
settings. Our results extend those of Hawkins and
Lidsey [Phys. Rev. D {\bf 66}, 023523 (2002)], by including
the non-flat case. The Ermakov-Pinney methodology is developed
in a general form, allowing through the converse results presented
herein to use it as a tool for constructing new solutions to 
the original equations. As an example of this type a special
blowup solution recently obtained in Christodoulakis {\it et al.}
[gr-qc/0302120] is retrieved. Additional solutions of the 3+1
dimensional gravity coupled with the scalar field are also
obtained.
To illustrate the generality of the approach, we extend it to
the anisotropic case of Bianchi types I and V and present
some related open problems.
\end{abstract}





\section{Introduction}

In the past few years, the ekpyrotic scenario has been
proposed as an alternative to the standard inflationary
cosmology \cite{lid8}. In this interpretation, the big 
bang is viewed as the collision of two domain walls or
branes described by Einstein's equations coupled to 
a scalar field. The scalar in this case parametrizes
the separation between the branes. Hence, in this 
dynamical setup, it is of interest to understand 
the coupling of scalar fields to gravity. 

This role has been investigated rather extensively 
in 2+1 dimensional setups; see e.g., the earlier
works of \cite{burd,cruz,floyd}. More recently the
3+1 dimensional case has become of interest; see e.g.,
\cite{hawk,our,campo,grqc}.

These studies have in fact motivated the present work.
In particular, in a significant recent paper \cite{hawk}, Hawkins
and Lidsey have proposed a connection between the 
so-called Ermakov-Pinney (EP) equation and flat,
3+1 dimensional, scalar field cosmologies. This is of
particular interest since the EP equation is a very
special, linearizable, nonlinear ordinary differential equation (ODE)
that can be solved exactly if the underlying linear
Schr{\"o}dinger equation can be solved.  We note in passing
that one of the authors of \cite{hawk} went on to use this
approach to illustrate analogies between scalar field cosmological
models and the dynamics of moments of the wavefunction of Bose-Einstein
condensates \cite{hawk1}. 
This point further highlights the importance of this ODE
that has been recurring in a variety of different areas
as diverse as nonlinear optics \cite{nlo}, elasticity \cite{nle}, 
quantum field theory \cite{qft} or molecular physics \cite{mol}. 
For a recent review of the EP equation and its applications, see
e.g., \cite{espin}.

One of the purposes of the present short communication is to demonstrate
how the methodology developed in \cite{hawk} can be generalized
in the case of non-zero curvature. We will also explicitly provide
a converse result according to which, given a solution of the
EP equation, the solution of a corresponding Einstein equation,
coupled to a scalar field, can be derived. The use of the converse
result will be explicitly demonstrated in a special case, that
of a linear Klein-Gordon equation for the scalar field. The
special solution for this case provided recently in \cite{our}
will be retrieved. More general solutions will also be constructed.

As our gravitational model, we will use the 3-dimensional 
Friedmann-Robertson-Walker (FRW) metric.
While other models such as the Brans-Dicke metric, see e.g., \cite{ricardo},
are also popular, the FRW metric is widely accepted  
in the early universe scenario that is of primary interest herein
(see the references mentioned above).
The relevant line element
(in co-moving coordinates and with a ``cosmological'' time
choice) will thus read:
\begin{equation}
ds^2=-dt^2+a^2(t) \left( \frac{1}{1-c r^2} dr^2 + r^2 d\theta^2 +
r^2 \sin^2 \theta d \phi^2 \right)
\label{ceq1}
\end{equation}
In the metric of Eq. (\ref{ceq1}), $a$ is the scale factor
while $c$ describes the curvature of the spatial slice and can be
normalized to the values $-1,0,1$ in the hyperbolic, flat and
elliptic case respectively. 

However, the above setting  seems to restrict our considerations
to an isotropic scenario. To illustrate the generality of the 
EP reduction and the usefulness of the corresponding technology,
we also examine in the present work, an anisotropic case, namely
the one of Bianchi types I and V. In the latter, spatially homogeneous,
yet anisotropic geometries, the line element is given by:
\begin{eqnarray}
ds^{2}=-N(t)^{2}dt^{2}+\gamma_{\alpha\beta}(t)\sigma^{\alpha}_{i}(x)
\sigma^{\beta}_{j}(x)dx^{i}dx^{j}
\label{ceqn1}
\end{eqnarray}
in the time gauge $N(t)=\sqrt{\gamma(t)}$; $\sigma^{\alpha}_{i}$ are
the invariant basis one-forms of the homegeneous surfaces of simultaneity
$\Sigma_t$, while $\gamma_{\alpha\beta}$ are the scale factors that
constitute the (in principle) 
dynamical variables of the metric. We restrict ourselves
in the present study to these anisotropic models, as these have been
argued to have the desirable property of 
isotropizing at arbitrarily long times \cite{collins}.  Additionally,
type V is the simplest Bianchi model that admits velocities or
tilted sources, hence it is natural to consider it as the simplest extension
that would allow the universe to choose a reference frame at the
exit from inflation (given that the de Sitter metric does not have
a preferred frame).

Our presentation will proceed as follows: in section 2 we will provide
the general EP methodology for the FRW system in the presence of a scalar.
In section 3, we will apply these results in the special case of
a massive scalar. In section 4 we will give additional special 
solutions to these equations, while in section 5, we will 
apply the method to the anisotropic case of Bianchi types I and V.
Finally, in section 6, we will summarize
our findings and present our conclusions.

\section{3+1 Isotropic Scalar Field with Curvature: the EP reduction}

We will follow the notation of \cite{hawk} in what follows. 
In particular, in this setting, the Einstein equations of gravity
and the Klein-Gordon equation for the scalar field can be
respectively written as
\begin{eqnarray}
H^2+\frac{c}{a^2}=\frac{\kappa^2}{3} \left[\frac{1}{2} \dot{\phi}^2
+ V(\phi) + \frac{D}{a^n} \right]
\label{ceq2}
\\
\ddot{\phi} + 3 H \dot{\phi} + \frac{d V}{d \phi}=0.
\label{ceq3}
\end{eqnarray}
$H=\dot{a}/a$ represents the Hubble parameter. 
The first two terms in the bracketed expression of  
Eq. (\ref{ceq2}) denote the energy density of the scalar
field with potential $V(\phi)$. The last term is the 
density of matter for the barotropic fluid with
equation of state $p_{mat}=(n-3) \rho_{mat}/3$; 
$D \geq 0$, $0 \leq n \leq 6$,
$\kappa^2=8 \pi/m_P^2$, where $m_P$ is the 
Planck mass. Finally, the dots will be used for
differentiation with respect to the cosmological or coordinate time $t$.
Notice that with respect to the corresponding equations
(2)-(3) of \cite{hawk}, the former is augmented by
the spatial curvature term (recall that $c$ can take the values
$-1, 0$ and $1$ depending on the curvature 
of the hypersurface $t=$const.) in the left hand side.

Now, by using a further differentiation of Eq. (\ref{ceq2})
and the substitution $b=a^{n/2}$ (cf. with Eq. (8) of
\cite{hawk}), we obtain:
\begin{eqnarray}
\frac{2}{n} \left(\frac{\ddot{b}}{b} - \frac{\dot{b}^2}{b^2} \right) - 
\frac{c}{b^{4/n}} = -\frac{\kappa^2}{2} \left[\dot{\phi}^2 + \frac{n D}{3 b^2}
\right].
\label{ceq4}
\end{eqnarray}
If we now define a new comoving time $\tau$ such that $\dot{\tau}=b$,
then Eq. (\ref{ceq4}) becomes
\begin{eqnarray}
\frac{d^2 b}{d \tau^2} + \frac{\kappa^2 n}{4} 
\left(\frac{d \phi}{d \tau} \right)^2 b= -\frac{\kappa^2 n^2 D}{12 b^3}
+ \frac{n c}{2 b^{\frac{4+n}{n}}},
\label{ceq5}
\end{eqnarray}
where the chain rule has been used and the functions in eq. (5) have been
assumed as
$b=b(t(\tau))$ and $\phi=\phi(t(\tau))$.
This equation can be compared with Eqs. (10)-(11) of \cite{hawk}, with
the difference being evident in the inclusion of the curvature term
(the last term on the right hand side of Eq. (\ref{ceq5})).

This prompts us to (briefly) discuss the EP equation which naturally arises
in this context not only in the flat case of $c=0$ examined in 
\cite{hawk}, but also in the case of $n=2$.
The latter is a remarkable
example of a nonlinear yet integrable 
ordinary differential equation (ODE) of the form:
\begin{eqnarray}
Y''+ Q(\tau) Y= \frac{\lambda}{Y^3}.
\label{oeq0}
\end{eqnarray}
The particularly appealing feature of this
nonlinear ODE is that its 
general solution can be obtained, provided that one is able
to solve the linear Schr{\"o}dinger (LS) problem $Y'' + Q(\tau) Y=0$.
For details on the properties of the EP equation, 
the interested reader is referred to
\cite{hawk,espin} and references therein. Here we just mention its basic
superposition principle property. Namely, if the linearly independent
solutions of the LS equation are $Y_1(\tau)$ and $Y_2(\tau)$, then the
most general possible solution of the EP equation is given by
\begin{eqnarray}
Y(\tau)=\left( A Y_1^2 + B Y_2^2 + 2 C Y_1 Y_2 \right)^{1/2}
\label{oeq8a}
\end{eqnarray}
where $A, B$ and $C$ are constants 
connected through
\begin{eqnarray}
A B - C^2= \frac{\lambda}{W^2}
\label{oeq8b}
\end{eqnarray}
and the Wronskian $W= Y_1 Y_2' - Y_2 Y_1'$.

It is of particular interest to note that the result of Eq.
(\ref{ceq5}) can be used conversely for constructing solutions
of scalar field cosmologies, using the EP equation structure
and solutions. The converse result can be proved in the
following form:
Given $Q$ and $\lambda=-\kappa^2 n^2 D/12 <0$, let $Y>0$ be
a solution of 
\begin{eqnarray}
\frac{d^2 Y}{d \tau^2} + Q Y= \frac{\lambda}{Y^3} + \frac{n c}{2 
Y^{1+\frac{4}{n}}}
\label{ceq6}
\end{eqnarray}
Define a new time coordinate $t$ such that $\dot{\tau}=Y(\tau(t))$ and
$a=Y^{2/n}$, as well as a new field $\phi$ satisfying:
\begin{eqnarray}
\frac{n \kappa^2}{4} \left(\frac{d \phi}{d \tau} \right)^2= Q,
\label{ceq7}
\end{eqnarray}
with $Q \neq 0$.
Finally, define a potential: 
\begin{eqnarray}
V(\phi)=\frac{12}{ \kappa^2 n^2} \left(\frac{d Y}{d \tau} \right)^2
-\frac{Y^2}{2} \left(\frac{d \phi}{d \tau} \right)^2 -\frac{D}{Y^2}
+ \frac{3 c}{\kappa^2 Y^{4/n}}.
\label{ceq8}
\end{eqnarray}
Then, if we consider the triplet $(a(\tau(t)),\phi(\tau(t)),V(\phi))$,
the latter satisfies the Eqs. (\ref{ceq2})-(\ref{ceq3}).

Let us now give a number of specific examples, where this construction
scheme can be used to obtain solutions to the $3+1$ dimensional
scalar field cosmology equations.

\section{Applications}

\subsection{Flat FRW Metric and Massless Scalar}

In the absence of matter, we can set $D=0 \Rightarrow \lambda=0$, 
and consider for convenience the case of $n=2$. For $\lambda=0$,
and for the flat FRW metric e.g., for $c=0$, Eq. (\ref{ceq6})
becomes
\begin{eqnarray}
\frac{d^2 Y}{d \tau^2} + Q Y= 0
\label{ceq9}
\end{eqnarray}
Assuming the general case $V(\phi)=m^2 \phi^2/2$, the particular
scenario of a massless scalar yields $V(\phi)=0$, hence
Eq. (\ref{ceq8}) becomes:
\begin{eqnarray}
\frac{d^2 Y}{d \tau^2} + \frac{3}{Y} \left( \frac{d Y}{d \tau} \right)^2 = 0
\label{ceq10}
\end{eqnarray}
whose solution yields:
\begin{eqnarray}
Y(\tau)=A \tau^{1/4};
\label{ceq11}
\end{eqnarray}
hence from Eq. (\ref{ceq9}):
\begin{eqnarray}
Q(\tau)= \frac{3}{16} \frac{1}{\tau^2}.
\label{ceq12}
\end{eqnarray}
This, in turn, through the definition of $\dot{\tau}$ yields:
\begin{eqnarray}
\tau(t)=\left(\frac{3}{4}\right)^{\frac{4}{3}} A^{4/3} t^{\frac{4}{3}}
\label{ceq13}
\end{eqnarray}
and, thus, finally from Eq. (\ref{ceq7}):
\begin{eqnarray}
\phi(t)=\sqrt{\frac{2}{3}} \log(\frac{3 A}{4}) + \sqrt{\frac{2}{3}} \log(t-t_0).
\label{ceq14}
\end{eqnarray}
which is the same solution as obtained in Eq. (8) of \cite{our}.

\subsection{A Generalization: Non Flat FRW Metric Coupled with Scalar}

To demonstrate the generality of the technique also in non-flat
cases with $c \neq 0$, we examine the example of $n=2$, $\kappa=1$
and $D > c/3$ (which implies that $\lambda+c <0$). 

In this case, Eq. (\ref{ceq6}) becomes:
\begin{eqnarray}
\frac{d^2 Y}{d \tau^2} + Q Y= \frac{\lambda + c}{Y^3}
\label{ceq15}
\end{eqnarray}
Motivated by cases in which we are able to solve the EP
(or equivalently the underlying linear Schr{\"o}dinger) 
equation exactly, we choose $Q(\tau)=3/(16 \tau^2)$ (cf. Eq. (\ref{ceq12})).
In this case, using Eq. (\ref{oeq8a}), we can find the
general solution to Eq. (\ref{ceq15}) as:
\begin{eqnarray}
Y(\tau)=\left( A \tau^{\frac{3}{4}} + B \tau^{\frac{1}{4}} + 2 C \tau
\right)^{\frac{1}{2}},
\label{ceq16}
\end{eqnarray}
where $A B - C^2=4 (\lambda+c)$. For convenience, we use our
freedom of coefficients to set $A=B=0$, hence $C=2 \sqrt{|\lambda+c|}$.
We thus obtain:
\begin{eqnarray}
Y(\tau)=(2 C \tau)^{\frac{1}{2}};
\label{ceq17}
\end{eqnarray}
From Eq. (\ref{ceq7}), we derive:
\begin{eqnarray}
\phi(\tau)=\frac{\sqrt{6}}{4} \log(\tau).
\label{ceq18}
\end{eqnarray}
From the definition of $\dot{\tau}$, it can be seen that:
\begin{eqnarray}
\tau=\frac{C}{2} (t-t_0)^2.
\label{ceq20}
\end{eqnarray}
Finally, from Eq. (\ref{ceq8}), it follows that:
\begin{eqnarray}
V(\phi)= \frac{S}{\tau} \equiv S e^{-\frac{4}{\sqrt{6}} \phi}
\label{ceq21}
\end{eqnarray}
where $S=9 c/8- D/(2 c) + 3/2$.

This result generalizes the corresponding result of Eq. (31) of \cite{hawk}
in the case in which curvature is present.


\subsection{An Example of Quadratic Scalar Fields}

As another example, we will use a case in which we use the
EP approach in a reverse engineering way. In particular, in
Eq. (\ref{ceq6}), we will postulate the solution and we will
obtain the potential that is compatible with this solution.
We assume that matter is absent, hence $D=\lambda=0$, 
for $n=4$ and $c \geq 0$. We now
demand that $Y(\tau)=2 B \tau$. Then 
\begin{eqnarray}
Q(\tau)=\frac{c}{4 B^3 \tau^3}.
\label{ceq22}
\end{eqnarray}
From the definition of $\dot{\tau}$, we obtain that
\begin{eqnarray}
\tau(t)=\frac{A^2}{2 B} e^{2 B t},
\label{ceq23}
\end{eqnarray}
where $A$ is (without loss of generality) a positive constant.
Then from Eq. (\ref{ceq7}), we obtain:
\begin{eqnarray}
\phi(t)=-\frac{\sqrt{2 c}}{\kappa A B} e^{- B t} + \alpha,
\label{ceq24}
\end{eqnarray}
where $\alpha$ is an arbitrary constant, while
the potential
\begin{eqnarray}
V(\phi)=\frac{3 B^2}{\kappa^2} + B^2 (\phi-\alpha)^2
\label{ceq25}
\end{eqnarray}
is quadratic in $\phi$. Notice that in this example, the restriction
of $n=4$ can be lifted: in particular substituting $A \rightarrow
A^{n/4}$ and $B \rightarrow B n/4$, we can perform the same 
calculation for any power $n$, but the expression for $V(\phi)$ is
functionally the same as the one of Eq. (\ref{ceq25}).

We remark here that the solution found in this subsection was previously
identified by means of a different (than the EP) approach in \cite{Ellis}
(cf. Eqs. (24)-(28) therein).

\subsection{An Example of Constant Scalar Fields}

In the same inverse procedure spirit, another convenient choice
of $Q(\tau)$ is $Q(\tau)=0$ (for $n=2$). Then the EP equation
has the straightforward general solution:
\begin{eqnarray}
Y(\tau)=\left(A \tau^2 + B + 2 C \tau \right)^{1/2},
\label{ceq26}
\end{eqnarray}
with
\begin{eqnarray}
A B - C^2= \lambda + c = -\frac{\kappa^2 D}{3} + c \equiv \tilde{\lambda}.
\label{ceq27}
\end{eqnarray}
Consequently, 
\begin{eqnarray}
\tau(t)=\frac{D}{4 A} \exp\left(A^{1/2} t\right) +
\frac{\tilde{\lambda}}{D A^{1/2}} \exp\left(-A^{1/2} t\right) -\frac{C}{A}. 
\label{ceq28}
\end{eqnarray}
It can then be immediately seen 
that the scalar $\phi$ and the potential $V$ are constants, while the
scale factor
\begin{eqnarray}
a(t)=a(0) \cosh \left(A^{1/2} t\right) + \sqrt{a(0)^2-
\frac{\tilde{\lambda}}{A}} \sinh \left(A^{1/2} t\right)
\label{ceq29}
\end{eqnarray}
with 
\begin{eqnarray}
a(0)= \frac{D}{4 A^{1/2}} + \frac{\tilde{\lambda}}{D A^{1/2}}. 
\label{ceq30}
\end{eqnarray}
It should be noted that this is a generalization of the solution
of Eq. (24) of \cite{hawk} in the spatially non-flat case.
Similar solutions have been obtained through direct calculations
(i.e., instead of the EP methodology) in \cite{cruz}.

\section{Anisotropic Generalizations: EP reduction for Bianchi
Types I and V}

\subsection{Bianchi Type I}

Since the anisotropic case was not previously discussed in 
the earlier work of \cite{cruz,floyd,hawk}, we discuss it here
in more detail.

In this case, the dynamical scale factors and the one-forms
in the metric of Eq. (\ref{ceqn1}) are given by:
\begin{displaymath}
\gamma_{\alpha\beta}(t)=\left(\begin{array}{ccc}
  A(t)^{2} & 0 & 0 \\
  0 & B(t)^{2} & 0 \\
  0 & 0 & \Gamma(t)^{2} \\
\end{array}\right), \quad
\sigma^{\alpha}_{i}(x)=\left(\begin{array}{ccc}
  1 & 0 & 0 \\
  0 & 1 & 0 \\
  0 & 0 & 1 \\
\end{array}\right)
\end{displaymath}

We can define the tensor: $F_{\mu\nu}=G_{\mu\nu}-8\pi T_{\mu\nu}$
where $G_{\mu\nu}=R_{\mu \nu}-\frac{1}{2} g_{\mu \nu} R$ is the Einstein 
tensor and $T_{\mu \nu}=\phi_{;\mu}\phi_{;\nu}
-\frac{1}{2}g_{\mu\nu}(\phi^{;\alpha}\phi_{;\alpha}+m^{2}\phi^{2})$
is the energy momentum tensor. Then, the quadratic constraint
is the equation $F_0^0=0$, the kinematic equation is given
by $F_1^1=0$, while the Klein-Gordon equation for the
field is given by $\phi^{;\mu}_{;\mu}-m^{2}\phi=0 \propto
T^{\mu\nu}_{;\nu}=0$.
Notice additionally, that the two integrals of the motion,
namely $I_1=F^{1}_{1}-F^{2}_{2}=0$ and $I_2=F^{1}_{1}-F^{3}_{3}=0$
yield $B(t)=A(t)e^{\kappa
t/2}$ and $\Gamma(t)=A(t)e^{\lambda t/2}$.

Solving the Klein-Gordon equation for $\phi''(t)$ and substituting
the result into $\partial_t F_0^0=0$ (as well as solving $F_0^0=0$
for $\phi$ and using the resulting expression in $\partial_t F_0^0=0$),
one is led  to a dynamical equation for the remaining scale factor $A(t)$
in the form:
\begin{equation}
\frac{\kappa \,\lambda }{4} +
  \frac{\kappa \,A'(t)}{A(t)} +
  \frac{\lambda \,A'(t)}{A(t)} +
  \frac{4\,{A'(t)}^2}{{A(t)}^2} -
  \frac{{\phi '(t)}^2}{2} - \frac{A''(t)}{A(t)}=0
\label{gop1}
\end{equation}
Using now: $A(t)=Y(t)^{2/n}$ and a change of variable
$\tau=\int^t \Omega(t') dt'$, we obtain:
\begin{equation}
\ddot{Y}(\tau)+\dot{Y}(\tau)\frac{\Omega'(t)}{\Omega(t)^{2}}-\frac{\dot{Y}(\tau)^{2}}
{Y(\tau)}\frac{(6+n)}{n}-(\kappa+\lambda)\frac{\dot{Y}(\tau)}{\Omega(t)}
-\frac{n\kappa\lambda}{8}\frac{Y(\tau)}{\Omega(t)^{2}}+
\frac{n Y(\tau)\dot{\phi}(\tau)^{2}}{4}=0
\label{gop2}
\end{equation}
Hence, a choice of time reparametrization according to:
\begin{equation}
\frac{\Omega '(t)}{\Omega (t)} =
  \kappa  + \lambda  +\frac{(6+n)}{n}\frac{Y'(t)}{Y(t)}
\label{gop3}
\end{equation}
(which leads to $\Omega(t)=\theta e^{(\kappa+\lambda)t} Y(t)^{(6+n)/n}$,
where $\theta>0$ is a constant of integration),
results in the form:
\begin{equation}
\ddot{Y}(\tau)+Q Y(\tau)=\frac{\Gamma}{Y(\tau)^{1+12/n}},
\quad Q=n\frac{\dot{\phi}(\tau)^{2}}{4}, \quad
\Gamma=\frac{n\kappa\lambda e^{-2(\kappa+\lambda)t}}{8\theta^{2}}.
\label{gop4}
\end{equation}
Eq. (\ref{gop4}) is of the form of Eq. (\ref{ceq5}) and becomes
an EP equation for the choice of $n=6$, for $\kappa=-\lambda$.


In the more general case, the problem becomes extremely complex
as $t$ depends on $\tau$ through $d \tau/dt = \Omega(t)$ and
$\Omega$ is itself a function of the solution. Typically, this
problem will not be analytically tractable (if combined, it leads
to an integro-differential equation for $Y(\tau)$).
However, this can
be used as an {\it inverse} problem: we can {\it postulate} 
$\tau(t)$, derive from it $\Omega(t)$ and $Y(t)$ and use
them in Eq. (\ref{gop4}) to derive the form of $\phi(t)$.

A simple example of the above methodology can be given as 
follows: let us assume that $\tau=e^{(\kappa + \lambda) t}/(\kappa+\lambda)$, 
then $\Omega(t)=e^{(\kappa + \lambda) t}$.
Hence, choosing $\theta=1$, for $n=6$ (the EP-like case
for  Eq. (\ref{gop4})) $Y(t)=Y(\tau)=1$.
Then, using the Eq.  (\ref{gop4})), we find that 
\begin{eqnarray}
\phi(\tau)=C + \sqrt{\frac{\kappa \lambda}{2 (\kappa + \lambda)^2}} \log(\tau)
\label{gopadd1}
\end{eqnarray}
and hence (e.g., choosing the integration constant $C=\log(\kappa+\lambda)
\sqrt{\frac{\kappa \lambda}{2 (\kappa + \lambda)^2}}$), $\phi(t)=
\sqrt{\frac{\kappa \lambda}{2}} t$.

\subsection{Bianchi Type V}

In this case, the scale factors and one-forms are, in turn, given
by:
\begin{displaymath}
\gamma_{\alpha\beta}(t)=\left(\begin{array}{ccc}
  B(t)^{2} & 0 & 0 \\
  0 & \Gamma(t)^{2} & 0 \\
  0 & 0 & A(t)^{2} \\
\end{array}\right), \quad
\sigma^{\alpha}_{i}(x)=\left(\begin{array}{ccc}
  0 & e^{-x} & 0 \\
  0 & 0 & e^{-x} \\
  1 & 0 & 0 \\
\end{array}\right).
\end{displaymath}

In type V, there exist three integrals of the motion, namely:
$I_1=F^{1}_{1}-F^{2}_{2}=0$, $I_2=F^{1}_{1}-F^{3}_{3}=0$ and
$I_3=F^{1}_{0}=0$, which yield: $B(t)=A(t)e^{\kappa t/2}$
as well as: $\Gamma(t)=A(t)e^{-\kappa
t/2}$. 

Once again following the same path as before and substituting
into $\partial_t F_0^0=0$, we obtain a single ODE for $A(t)$
of the form:
\begin{eqnarray}
-6\,{\kappa }^2\,A(t)\,A'(t) - 24\,{A(t)}^5\,A'(t) +
  \frac{96\,{A'(t)}^3}{A(t)} - \nonumber \\
  12\,A(t)\,A'(t)\,{\Phi '(t)}^2 - 24\,A'(t)\,A''(t)=0
\label{gop5}
\end{eqnarray}
Through a similar motivation as in the previous subsection,
we use the change of variables: $A(t)=Y(t)^{2/n}$, alongside
the time reparametrization $\tau=\int^t \Omega(t') dt'$,
with $\Omega(t)=\theta Y^{(6+n)/n}$. 
This results in the final form:
\begin{eqnarray}
\ddot{Y}(\tau)+Q Y(\tau)=-\frac{n\kappa^{2}}{8\theta^{2}}\frac{1}{Y(\tau)^{1+12/n}}
-\frac{n}{2\theta^{2}}\frac{1}{Y(\tau)^{1+4/n}}, 
\label{gop6}
\end{eqnarray}
where $
Q=n\frac{\dot{\phi}(\tau)^{2}}{4}$, which is again a generalized form
of an EP equation. This statement is made in the sense that while there is 
no choice of the
exponent that yields a mere inverse cubic dependence on $Y(t)$
in the right hand side of Eq. (\ref{gop6}), 
the typical scenario involves inverse power dependence in a manner
similar to the EP equation. 


While the resulting Eq. (\ref{gop6}) is not directly of the form of the
EP equation, its dynamics can be controllably tuned to be close to those
of the EP solutions. In particular,
it is clear that the asymptotic behavior of the
equation will be similar to that of $\ddot{Y}(\tau) +Q Y(\tau)=-
\frac{n\kappa^{2}}{8\theta^{2}}\frac{1}{Y(\tau)^{1+12/n}}$ 
for small initial data, while it will
be close to that of 
$\ddot{Y}(\tau) +Q Y(\tau)=-\frac{n}{2\theta^{2}}\frac{1}{Y(\tau)^{1+4/n}}$ 
for the case of large initial data. Hence, the choice of $n=6$ for 
small initial data and the one of $n=2$ for large initial data yields
EP behavior (this point has also been verified in numerical investigations
of the equation not shown here). In the intermediate regime where there
is competition between the two terms the EP results will not be immediately
applicable and one should resort to numerical simulations 
of Eq. (\ref{gop6}).


\section{Conclusions}

In this  short communication, we have generalized the earlier work
of \cite{hawk} and of \cite{floyd} in the direction of 
obtaining solutions to $3+1$ dimensional cosmological models for the 
Friedmann-Robertson-Walker case in a systematic way. The method
re-casts the relevant ordinary differential equations into one
of the Ermakov-Pinney type which is explicitly solvable.
From the solutions of the resulting EP equation one can
re-construct the solutions to the original cosmological model 
in a step-by-step inverse process that has been detailed. 
The present study generalizes that of \cite{hawk} in that
cases of non-zero curvature of the FRW metric can be also
considered. It also extends the results of \cite{floyd} in
the $3+1$ dimensional context. The method can be used to
derive a variety of solutions in the latter context, including
ones considered previously in \cite{burd,cruz,our}.


On the other hand, we have also illustrated the generality of 
our method, by exploring the potential of such Ermakov-Pinney
reductions to systems with anisotropy. In particular, we have
demonstrated that the case examples of Bianchi types I and V
can be reduced to Ermakov-Pinney-like equations. These Bianchi
types were chosen as prototypical examples where a velocity/tilt
can be included and as examples of Bianchi types that may eventually
isotropize.
The EP reduction of these cases highlights the generality and usefulness of the 
procedure. However, by the same token and since the EP is the
only one of these inverse power, nonlinear ordinary differential
equations for which an explicit solution exists, it underlines
the importance of understanding the behavior of such classes
of equations. In particular, examining numerically their temporal
evolution for a number of physically motivated cases would be 
a natural next step.
Such studies are currently 
in progress and will be reported in a future work.


G.O. Papadopoulos is currently a scholar of the Greek State
Scholarships Foundation (I.K.Y.) and acknowledges the relevant
financial support.
T Christodoulakis and G.O. Papadopoulos, acknowledge support by
the University of Athens, Special Account for the Research
Grant-No. 70/4/5000
This work was also partially supported by the Eppley 
Foundation for Research, NSF-CAREER and NSF-DMS-0204585 (PGK).
We are thankful to J.E. Lidsey (private 
communication with T.C.) for useful discussions and
for pointing out to us the connection between the results of
Section 3.3 and those of \cite{Ellis}.



\end{document}